\documentclass[12pt,aps,prd,preprint,tightenlines,superscriptaddress,showpacs,nofootinbib]
{revtex4-1}
\usepackage{epsfig}
\usepackage{epstopdf}
\usepackage{amssymb,amsmath}

\newcommand{\bea}{\begin{eqnarray}}
\newcommand{\eea}{\end{eqnarray}}

 \makeatletter
\def\alt{\mathrel{\mathpalette\gl@align<}}
\def\agt{\mathrel{\mathpalette\gl@align>}}
\def\gl@align#1#2{\lower.6ex\vbox{\baselineskip\z@skip\lineskip\z@
\ialign{$\m@th#1\hfil##\hfil$\crcr#2\crcr\sim\crcr}}} \makeatother

\begin{document}
\vspace*{1.0cm}

\begin{center}
\baselineskip 20pt 
{\Large\bf 
A natural $Z^\prime$-portal Majorana dark matter\\
 in alternative U(1) extended Standard Model
}
\vspace{1cm}

{\large 
Nobuchika Okada$^{~a}$\footnote{okadan@ua.edu},  
Satomi Okada$^{~a}$\footnote{satomi.okada@ua.edu}
and Digesh Raut$^{~b}$\footnote{draut@udel.edu}
}
\vspace{.5cm}

{\baselineskip 20pt \it
$^a$Department of Physics and Astronomy, University of Alabama, Tuscaloosa, AL35487, USA\\
$^b$Bartol Research Institute, Department of Physics and Astronomy, \\
University of Delaware, Newark, DE 19716, USA
}

\vspace{.5cm}

\vspace{1.5cm} {\bf Abstract}
\end{center}

We consider a non-exotic gauged U(1)$_X$ extension of the Standard Model (SM), 
  where the U(1)$_X$ charge of a SM field is given by a linear combination of its hypercharge 
  and Baryon-minus-Lepton ($B-L$) number. 
All the gauge and mixed gauge-gravitational anomalies are canceled in this model with the introduction of 
  three right-handed neutrinos (RHNs). 
Unlike the conventional minimal U(1)$_X$ model, where a universal U(1)$_X$ charge of $-1$ is assigned to 
  three RHNs, we consider an alternative charge assignment, namely, 
  two RHNs ($N_R^{1,2}$) have U(1)$_X$ charge $-4$  while one RHN ($N_R$) has a $+5$ charge.   
With a minimal extension of the Higgs sector, the three RHNs acquire their Majorana masses 
  associated with U(1)$_X$  symmetry breaking. 
While $N_R^{1,2}$ have Yukawa coupling with the SM lepton doublets 
  and play an essential role for the ``minimal seesaw'' mechanism, 
  $N_R$ is isolated from the SM particles due to its U(1)$_X$ charge and 
  hence it is a natural candidate for the dark matter (DM) without invoking additional symmetries. 
In this model context, we investigate the $Z^\prime$-portal RHN DM scenario, 
  where the RHN DM communicates with the SM particles through the U(1)$_X$ gauge boson ($Z^\prime$ boson). 
We identify a narrow parameter space by combining the constraints from 
  the observed DM relic abundance, the results of the search for a $Z^\prime$ boson resonance 
  at the Large Hadron Collider Run-2, 
  and the gauge coupling perturbativity up to the Planck/Grand Unification scale.  
A special choice of U(1)$_X$ charges for the SM fields allows us to extend the model 
  to SU(5)$\times$U(1)$_X$ grand unification. 
In this scenario, the model parameter space is more severely constrained, 
  which will be explored at future high energy collider experiments.

\thispagestyle{empty}

\newpage

\addtocounter{page}{-1}

\baselineskip 18pt

\section{Introduction} 
\label{sec:1}

Despite its tremendous success in describing various elementary particle phenomena, 
  the Standard Model (SM) cannot account for two major experimental results, 
  namely, the existence of dark matter (DM) in the Universe and the neutrino oscillation phenomena 
  resulting in neutrinos' tiny masses and substantial mixings among different flavors.
Type-I seesaw \cite{seesaw} is probably the simplest possibility to naturally generate tiny neutrino masses, 
  where heavy Majorana right-handed neutrinos (RHNs) singlet under the SM gauge group play the crucial role. 
These RHNs are naturally incorporated into the minimal gauged $B-L$ extension of the SM \cite{MBL}, 
  where the global $B-L$ (Baryon number minus Lepton number) symmetry of the SM is gauged. 
In the model, all the gauge and mixed gauge-gravitational anomalies are canceled out 
  in the presence of three RHNs. 
Majorana masses for the RHNs are generated associated with the spontaneous U(1)$_X$ symmetry breaking,  
  and the type-I seesaw mechanism works after the electroweak symmetry breaking.

A concise way to incorporate a DM candidate into the minimal $B-L$ has been proposed in Ref.~\cite{OS},  
  where a $Z_2$ symmetry is introduced and an odd-parity is assigned to one RHN 
  while all the other fields are parity-even. 
Requiring the Lagrangian to be $Z_2$-even, the parity-odd RHN is stable and hence a unique DM candidate in the model. 
The essential point is that the $Z_2$-parity devides three RHNs into two parity-even RHNs and the RHN DM, 
  so that the seesaw mechanism is realized with only the two RHNs. 
This framework is the so-called ``minimal seesaw'' \cite{MinSeesaw}, which has a sufficient number of free parameters 
  to reproduce the neutrino oscillation data while predicting one massless light neutrino eigenstate. 
The phenomenology of the RHN DM in the minimal $B-L$ model has been 
  extensively studied \cite{OS, RHNDM1, RHNDM2, OO1, Klasen:2016qux, SO, Escudero:2018fwn}. 
It is particularly interesting that the RHN DM communicates with the SM particles through 
  the $Z^\prime$ boson ($Z^\prime$-portal RHN DM). 
As has been shown in Ref.~\cite{OO1}, the cosmological constraint and
  the results from the search for a $Z^\prime$ boson resonance at the Large Hadron Collider (LHC) Run-2   
  are complementary to narrow down the model parameter space. 
It is known that the minimal $B-L$ model is generalized to the minimal U(1)$_X$ model \cite{MinU1X} 
  with the same particle content except for the U(1)$_X$ charge assignment \cite{U1X}: 
  the U(1)$_X$ charge of a SM field is defined as a linear combination of its hypercharge and $B-L$ charge. 
The phenomenology of the $Z^\prime$-portal RHN DM in the context of the minimal U(1)$_X$ model 
   has been studied in Ref.~\cite{OO2} (see also \cite{Oda:2017kwl}) to identify the allowed parameter region 
   from the cosmological and collider phenomenology constraints.    
Furthermore, an extension of the U(1)$_X$ model to SU(5)$\times$U(1)$_X$ grand unification 
  has been proposed in Ref.~\cite{OOR}.

In the minimal $B-L$ model or the U(1)$_X$ model discussed above,  
  flavor universal U(1)$_X$ charges are assigned to three RHNs. 
However, there is another charge assignment to make the model anomaly-free \cite{AltU1X}. 
In this alternative charge assignment, two RHNs have U(1)$_X$ charge $-4$ 
  while a $+5$ charge is assigned to one RHN. 
Interestingly, the three RHNs are divided into two RHNs and one RHN by their U(1)$_X$ charges. 
As we will discuss in the next section, in a minimal extension of the Higgs sector, 
  the two RHNs are involved in the minimal seesaw mechanism, while the RHN 
  with $+5$ charge cannot have coupling with the SM particles because of its U(1)$_X$ charge. 
Hence, this RHN with $+5$ charge is a natural DM candidate without introducing any other symmetry like $Z_2$ symmetry. 
In this paper we investigate the $Z^\prime$-portal RHN DM scenario in the U(1)$_X$ extended SM 
  with the alternative U(1)$_X$ charge assignment for the RHNs 
  (see, for example, Ref.~\cite{Ma:2014qra} for an attempt to implement the inverse seesaw with the alternative charge assignment).    
Considering the cosmological constraint, the LHC Run-2 results for the $Z^\prime$ boson resonance search, 
  and the gauge coupling perturbativity constraint, 
  we identify an allowed model parameter space. 
We also discuss an extension of the model to SU(5)$\times$U(1)$_X$ grand unification 
  and phenomenological constraints of the model.

This paper is organized as follows: 
In the next section, we introduce the U(1)$_X$ model with the alternative charge assignment for RHNs and a minimal Higgs sector. 
In Sec.~\ref{sec:3}, we analyze the DM relic abundance and constrain on the model parameter region 
  so as to reproduce the observed DM relic abundance. 
In Sec.~\ref{sec:4}, we consider the LHC Run-2 results by the ATLAS and the CMS collaborations 
   from the search for a narrow resonance with  dilepton final states. 
We interpret the current LHC results into the constraints on the $Z^\prime$ boson production process 
   in our U(1)$_X$ extended SM. 
We also consider the gauge coupling perturbativty bound as well as the LEP constraints from the search 
   for effective 4-Fermi interactions. 
Combining all the constraints, we identify an allowed parameter region. 
We further discuss an extension of the model to SU(5)$\times$U(1)$_X$ grand unification  in Sec.~\ref{sec:5}. 
The last section is devoted to conclusions and discussions.

\section{The U(1)$_X$ extended SM with Alternative Charge Assignment}
\label{sec:2}
\begin{table}[t]
\begin{center}
\begin{tabular}{c|ccc|c|c}
      &  SU(3)$_C$  & SU(2)$_L$ & U(1)$_Y$ & U(1)$_X$ \\ 
\hline
$q^{i}_{L}$ & {\bf 3 }    &  {\bf 2}         & $ 1/6$       & $(1/6) x_{H} + (1/3)  $ \\
$u^{i}_{R}$ & {\bf 3 }    &  {\bf 1}         & $ 2/3$       & $(2/3) x_{H} + (1/3)  $ \\
$d^{i}_{R}$ & {\bf 3 }    &  {\bf 1}         & $-1/3$       & $(-1/3) x_{H} + (1/3) $\\
\hline
$\ell^{i}_{L}$ & {\bf 1 }    &  {\bf 2}         & $-1/2$       & $(-1/2) x_{H} +(-1)   $ \\
$e^{i}_{R}$    & {\bf 1 }    &  {\bf 1}         & $-1$                   & $ (-1) x_{H} +(-1) $ \\
\hline
$H$            & {\bf 1 }    &  {\bf 2}         & $- 1/2$       & $(-1/2) x_{H}$ \\  
\hline
\hline
$N^{1}_{R}$    & {\bf 1 }    &  {\bf 1}         &$0$                    & $- 4 $ \\ 
$N^{2}_{R}$    & {\bf 1 }    &  {\bf 1}         &$0$                    & $- 4 $ \\
$N_{R}$    & {\bf 1 }    &  {\bf 1}         &$0$                           & $+ 5 $   \\
\hline
$H_\nu$            & {\bf 1 }       &  {\bf 2}       &$ -\frac{1}{2}$                  & $ (-1/2) x_{H}+3 $  \\ 
$\Phi_A$            & {\bf 1 }       &  {\bf 1}       &$ 0$                  & $ +8  $  \\ 
$\Phi_B$            & {\bf 1 }       &  {\bf 1}       &$ 0$                  & $ -10 $  \\ 
\end{tabular}
\end{center}
\caption{
The particle content of the U(1)$_X$ extended SM with an alternative charge assignment. 
In addition to the three generations of SM particles ($i=1,2,3$), 
  the particle content includes three RHNs ($N_R^{1,2}$ and $N_R$) and three Higgs fields ($H_\nu, \Phi_{A,B}$).  
The U(1)$_X$ charge of a SM field is defined as a linear combination of the SM U(1)$_Y$ and the U(1)$_{B-L}$ charges 
  with one real parameter $x_H$.  
The model is anomaly free in the presence of the three RHNs with their assigned U(1)$_X$ charges. 
}
\label{table1}
\end{table}

We consider a U(1)$_X$ extended SM, where the U(1)$_X$ charge of a SM field is defined as a linear combination 
  of its hypercharge ($Q_Y$) and $B-L$ charge ($Q_{B-L}$), $Q_X = Q_Y x_H + Q_{B-L}$.
Here, a real parameter $x_H$ parameterizes an ``angle'' between the U(1)$_Y$ and the U(1)$_{B-L}$ directions.
Unlike the conventional case with a generation independent U(1)$_X$ charge assignment for three RHNs, 
   we consider an alternative charge assignment, namely, a  U(1)$_X$ charge $-4$ is assigned to two RHNs ($N_R^{1,2}$)  
   while a U(1)$_X$ charge $-5$ is assigned for the third RHN ($N_R$) \cite{AltU1X}.
The cancellation of all the gauge and mixed gauge-gravitational anomalies is also achieved by this charge assignment. 
In this ``alternative U(1)$_X$ model'', we introduce a minimal Higgs sector with a new Higgs doublet $H_\nu$ 
  and U(1)$_X$ Higgs scalars  $\Phi_{A,B}$. 
The particle content is listed in Table~\ref{table1}.\footnote{
With this minimal scalar particle content, we will have Nambu-Goldstone modes more than 
  those eaten by the weak and $Z^\prime$ bosons since mixing mass terms for the scalars
  are forbidden by the gauge symmetry.     
In particular, electrically-charged components in $H_\nu$ must be massive for our model
  to be phenomenologically viable. To eliminate such phenomenologically dangerous massless
modes, we need additional scalar fields. 
Since adding new scalars to the model has nothing to do with the gauge anomalies,
 it is straightforward to ameliorate the problem with additional scalars and there are many
possibilities. Among them, a simple way to generate a mass mixing between $H$ and $H_\nu$ 
 will be discussed in the last paragraph in Sec.~\ref{sec:Conclusions}.
}

In addition to the SM, we introduce Yukawa couplings involving new fields: 
\bea
\mathcal{L} _{Y} = 
-\sum_{i=1}^{3} \sum_{j=1}^{2} Y_{D}^{ij} \overline{\ell_{L}^{i}} H_\nu N_{R}^{j}
-\frac{1}{2} \sum_{k=1}^{2} Y_{N}^{k} \Phi_A \overline{N_{R}^{k \,c}} N_{R}^{k}
-\frac{1}{2} Y_{N}^{3} \Phi_B \overline{N_{R}^{~c}} N_{R}+ \rm{h. c.}.
\label{ExoticYukawa}
\eea
Note that due to the gauge invariance, 
   two RHNs ($N_R^{1,2}$) have the Dirac Yukawa couplings with the SM lepton doublets 
   as well as  the Majorana Yukawa couplings, 
   while  the RHN $N_R$ has only the Majorana Yukawa coupling. 
In the scalar sector, we assume a suitable Higgs potential to yield vacuum expectation values (VEVs) for $H$, $H_\nu$, $\Phi_{A}$, and $\Phi_B$: 
\bea
  \langle H \rangle =  \left(  \begin{array}{c}  
    \frac{1}{\sqrt{2}}v_h \\
    0 \end{array}
\right),   \;  \;  \; \; 
\langle H_\nu \rangle =  \left(  \begin{array}{c}  
    \frac{1}{\sqrt{2}} v_\nu\\
    0 \end{array}
\right),  \;  \;  \; \; 
\langle \Phi_A \rangle =  \frac{v_{A}}{\sqrt{2}},  \;  \;  \; \; 
\langle \Phi_B \rangle =  \frac{v_{B}}{\sqrt{2}}, 
\eea   
where we require $v_h^2 + v_\nu^2 = (246 \;  {\rm GeV})^2$ for the electroweak symmetry breaking. 
After the U(1)$_X$ and SM gauge symmetries are spontaneously broken, 
  the Majorana mass terms for the RHNs and the mass of the U(1)$_X$ gauge boson ($Z^\prime$) are generated:  
\bea
 m_{N^{1,2}}&=&\frac{Y_N^{1,2}}{\sqrt{2}} v_A,  \; \;  \; \;  \; \;
 m_N=\frac{Y_N}{\sqrt{2}} v_B,  \nonumber \\ 
 m_{Z^\prime} &=& g_X \sqrt{64 v_{A}^2+ 100 v_{B}^2+  \frac{1}{4} x_H^2 v_h^2 + \left(-\frac{1}{2} x_H +3\right)^2 v_\nu^2} \nonumber \\
 &\simeq& g_X \sqrt{64 v_{A}^2+ 100 v_{B}^2} . 
\eea 
Here, we have used the LEP constraint: $v_A^2+v_B^2 \gg v_h^2 + v_\nu^2$ \cite{Carena:2004xs}. 
Similarly, the neutrino Dirac masses are given by 
\bea
 m_{D}^{ij}=\frac{Y_{D}^{ij}}{\sqrt{2}} v_\nu, 
\eea
and hence the minimal seesaw mechanism with only two RHNs ($N_R^{1,2}$) generates tiny masses for the SM neutrinos. 
Thanks to its U(1)$_X$ charge, $N_R$ has no direct coupling with the SM fields, and hence it is naturally a DM candidate. 
This is in a sharp contrast with the minimal U(1)$_X$ model with a RHN DM \cite{OS, OO2},  
  where the introduction of $Z_2$ symmetry is essential to stabilize a $Z_2$-odd RHN DM 
  with the conventional U(1)$_X$ charge assignment.

\section{Cosmological bounds on $Z^\prime$-portal RHN DM.}
\label{sec:3}
Let us first consider the DM physics of our model. 
There are two ways for the RHN DM to interact with the SM particles. 
One is through the $Z^\prime$ boson interaction since all particles in our model have U(1)$_X$ charges 
  ($Z^\prime$-portal DM). 
The other is through the Higgs boson interactions. 
Since the Higgs field $\Phi_B$ generally has mixed quartic couplings with 
  the other Higgs fields in the Higgs potential, 
   the Higgs boson mass eigenstates after the gauge symmetry breaking 
   include the SM Higgs boson components. 
As a result, a pair of RHN DMs can communicate with the SM particles 
   through the Higgs bosons (Higgs-portal DM). 
Since the Higgs-portal DM scenario has already been extensively studied in the literature \cite{OS, RHNDM2}, 
   we assume that the mixing coupling between $\Phi_ B$ and $H$ is negligibly small 
   and focus on the $Z^\prime$-portal RHN DM scenario. 
As previously studied in Refs.~\cite{OO1, OO2}, 
   the DM physics and the $Z^\prime$ boson search at the LHC are complementary 
   to narrow down the model parameter space.

The relic abundance of the DM measured by the Planck satellite experiments 
  is given by $\Omega_{DM} h^2 = 0.1198\pm 0.0015$ \cite{PlanckDM} (at 68\% limit). 
In our analysis, we impose the constraint on the parameters so as to reproduce 
   the observed DM relic abundance, $0.1183 \leq  \Omega_{DM} h^2 \leq 0.1213$. 
To evaluate the DM relic abundance, we solve the Boltzmann equation given by \cite{KolbTurner}
\bea 
  \frac{dY}{dx}
  = - \frac{\langle \sigma v \rangle}{x^2}\frac{s (m_{DM})}{H(m_{DM})} \left( Y^2-Y_{EQ}^2 \right), 
\label{Boltmann}
\eea  
where $x=m_{DM}/T$ is a ``temperature'' normalized by the DM mass $m_{DM}$, 
  $\langle \sigma v \rangle$ is a thermally averaged DM annihilation cross section ($\sigma$) times relative velocity ($v$), 
  $H(m_{DM})$ is the Hubble parameter at $T=m_{DM}$, 
  $s(m_{DM})$ is the entropy density of the thermal plasma at $T=m_{DM}$,  
  $Y$ is the yield of the DM particle which is defined as a ratio of the DM number density to the entropy density, 
  and $Y_{EQ}$ is the yield of the DM in thermal equilibrium. 
Explicit forms for the quantities in the Boltzmann equation are as follows:   
\bea 
 H(m_{DM}) &=&  \sqrt{\frac{\pi^2}{90} g_\star} \frac{m_{DM}^2}{M_P}, \nonumber \\
 s(m_{DM}) &=& \frac{2  \pi^2}{45} g_\star m_{DM}^3,  \nonumber \\
 Y_{EQ}(x) &=&  \frac{g_{DM}}{2 \pi^2} \frac{x^2 m_{DM}^3}{s(m_{DM})} K_2(x),   
\eea
where $K_2 (x) $ is the modified Bessel function of the second kind, 
   $M_P=2.44 \times 10^{18}$ GeV is reduced Planck mass,
   $g_{DM}=2$ is the number of degrees of freedom for the Majorana RHN DM, 
   and $g_\star$ is the effective total number of degrees of freedom for the particles in thermal equilibrium 
   which we fix $g_\star=106.75$ for the SM particles. 
The thermal average of the DM annihilation cross section is given by the following integral expression: 
\bea 
\langle \sigma v \rangle =  \frac{g_{DM}^2}{64 \pi^4}
  \left(\frac{m_{DM}}{x}\right) \frac{1}{n_{EQ}^{2}}
  \int_{4 m_{DM}^2}^\infty  ds \; \hat{\sigma}(s) \sqrt{s} K_1 \left(\frac{x \sqrt{s}}{m_{DM}}\right),
\label{ThAvgSigma}
\eea
where $n_{EQ}=s(m_{DM}) Y_{EQ}/x^3$ is the DM number density, 
  $K_1$ is the modified Bessel function of the first kind, and 
  the reduced cross section $\hat{\sigma}(s)$ is defined as 
\bea
\hat{\sigma}(s)=2 (s- 4 m_{DM}^2) \sigma(s),
\eea
  with $\sigma(s)$ being the total annihilation cross section of the DM particle.

Through the $Z^\prime$ boson exchange, a pair of RHN DMs annihilates into SM fermion pairs 
  and pairs of the other RHNs if kinematically allowed.\footnote{  
We do not consider the the final state involving all exotic Higgs bosons, 
  assuming them to be heavier than the RHN DM.  
For $x_H \neq 0$, a pair of RHN DMs can annihilate into the SM $Z$ and Higgs bosons. We find the contribution of this process to the total annihilation cross section is negligibly small. When the $Z^\prime$ boson is lighter than the RHN DM, a pair of RHN DMs annihilates into a pair of $Z^\prime$ bosons. 
As we will discuss in the next section, the U(1)$_X$ gauge coupling is constrained to be very small from the LHC Run-2 results and the gauge coupling perturbativity. 
With such a small gauge coupling, the cross section of this process is too small to reproduce the observed DM relic density. }  
The RHN DM pair annihilation cross section for these processes are given by 
\bea 
 \sigma_{SM}(s) &=&\frac{25 \pi}{3}  \alpha_X^2  \frac{\sqrt{s (s-4 m_{DM}^2)}}
  {(s-m_{Z^\prime}^2)^2+m_{Z^\prime}^2 \Gamma_{Z^\prime}^2}  F(x_H),  \nonumber    \\
\sigma_{N^iN^i}(s) &=&
\frac{400\pi}{3}\alpha_X^2 \sqrt{\frac{s-4m_{N^i}^2}{s-4m_{DM}^2}} \frac{1} {(s-m_{Z^\prime}^2)^2+m_{Z^\prime}^2 \Gamma_{Z^\prime}^2}\nonumber \\ 
&\times & 
\frac{1}{s}\left((s-4m_{DM}^2)(s-4m_{{N^i}}^2)+ 12 \frac{m_{DM}^2 m_{N^i}^2}{m_{Z^\prime}^4} \left(s-m_{Z^\prime}^2\right)^2\right) 
 \theta(s-4 m_{N^i}^2), 
\label{DMSigma}
\eea 
where  
\bea 
  F(x_H)=13+ 16 x_H  + 10 x_H^2 = 10  \left( x_H+\frac{4}{5} \right)^2 +\frac{33}{5},  
\label{F}  
\eea
$\alpha_X=g_X^2/(4 \pi)$ is the U(1)$_X$ gauge coupling, 
   $\theta$ is the Heaviside step function, and $\Gamma_{Z'}$ is the total decay width of the  $Z^\prime$ boson given by 
\bea
\Gamma_{Z'} &=& 
 \frac{\alpha_X}{6} m_{Z^\prime} 
 \left[ F(x_H) + 5^2\left( 1-\frac{4 m_{DM}^2}{m_{Z^\prime}^2} \right)^{\frac{3}{2}} 
 \theta \left( \frac{m_{Z^\prime}^2}{m_{DM}^2} - 4 \right) \right. \nonumber \\ &+& \left. \sum_{i=1}^2 4^2\left( 1-\frac{4 m_{N^i}^2}{m_{Z^\prime}^2} \right)^{\frac{3}{2}} 
 \theta \left( \frac{m_{Z^\prime}^2}{m_{N^i}^2} - 4 \right)  \right]. 
\label{width}
\eea
In our analysis, we have neglected all SM fermion masses, since the RHN DM and $Z^\prime$ boson are  
  much heavier than the SM particles as we will see  below.  
Then, the total annihilation cross section of the RHN DM is given by
\bea
    \sigma(s) = \sigma_{SM}(s) +  \sum_{i=1}^2\sigma_{N^iN^i}(s). 
\eea 
By numerically solving the Boltzmann equation, we evaluate the relic abundance of the RHN DM at the present Universe by 
\bea 
  \Omega_{DM} h^2 =\frac{m_{DM} s_0 Y(\infty)} {\rho_c/h^2}, 
\eea 
where $s_0 = 2890$ cm$^{-3}$ is the entropy density of the present Universe, 
   and $\rho_c/h^2 =1.05 \times 10^{-5}$ GeV/cm$^3$ is the critical density.

\begin{figure}[t]
\begin{center}
\includegraphics[width=0.7\textwidth, height=7cm]{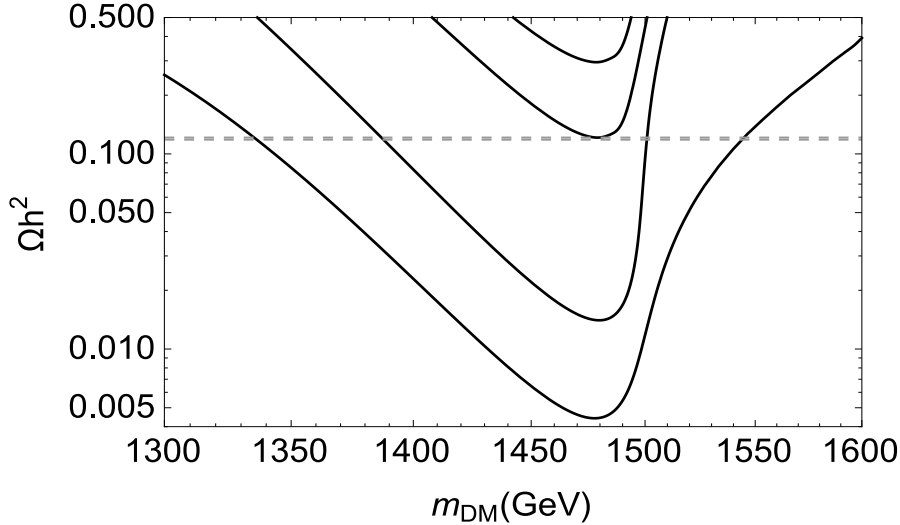} 
\end{center}
\caption{
The DM relic abundance as a function of the DM mass ($m_{DM}$) 
   for a fixed $m_{Z^\prime}=3$ TeV and $x_H = -0.8$. 
The solid line from top to bottom correspond to the resultant relic abundances  
    for $\alpha_X=2.0 \times 10^{-5}$,  $5.06\times 10^{-5}$,  $5.0\times 10^{-4}$, and $2.0\times 10^{-3}$, respectively.  
The two horizontal dashed lines indicate the range of the observed DM relic density,  $0.1183 \leq \Omega_{DM} h^2 \leq 0.1213$.
}
\label{fig:Omega}
\end{figure}

In our analysis, we set $m_N^{1,2} =  m_{Z^\prime}/4$, for simplicity.\footnote{
In Ref.~\cite{DOR}, a prospect of discovering the RHNs ($N^{1,2}$) at the High-Luminosity LHC 
  has been investigated in the same model context with the parameter choice of 
  $m_N^{1,2}  = m_{Z^\prime}/4 =750$ GeV. 
When we set $m_{Z^\prime}=3$ TeV in our analysis, we can combine our results of the present paper with those in Ref.~\cite{DOR}. 
}
The resultant DM relic abundance is controlled by four free parameters, 
   namely, $\alpha_X$, $m_{Z^\prime}$, $m_{DM}$, and $x_H$.  
In Fig.~\ref{fig:Omega}, we show the relic abundance as a function of $m_{DM}$ 
   for fixed $m_{Z^\prime}=3$ TeV and $x_H = -0.8$ as an example, 
   along with the observed DM relic abundance in the range of $0.1183 \leq \Omega_{DM} h^2 \leq 0.1213$ 
   (the region between two dashed lines).  
The solid lines from top to bottom are the relic abundance for fixed gauge coupling values,   
   for $\alpha_X=2.0 \times 10^{-5}$,  $5.06\times 10^{-5}$,  $5.0\times 10^{-4}$, and $2.0\times 10^{-3}$, respectively.    
From Fig.~\ref{fig:Omega} we have found a lower bound on $\alpha_X \gtrsim 5.06\times 10^{-5}$ 
  in order to reproduce the observed DM relic abundance. 
Fig.~\ref{fig:Omega} also indicates that the $Z^\prime$ boson resonance effect is crucial 
  in reproducing the observed DM relic abundance and hence, $m_{DM} \simeq m_{Z^\prime}/2$. 
As $\alpha_X$ is increased, the DM mass to reproduce the observed DM relic abundance is 
  going away from $m_{Z^\prime}/2$. 
However, as we will find in the following sections, there is an upper bound on $\alpha_X$ 
  from the LHC and the perturbativity constraints. 
Under the upper bound, we always find the DM mass is close to the $Z^\prime$ resonant point, 
  $m_{DM} \simeq m_{Z^\prime}/2$.

\begin{figure}[t]
\begin{center}
\includegraphics[width=0.7\textwidth, height=7cm]{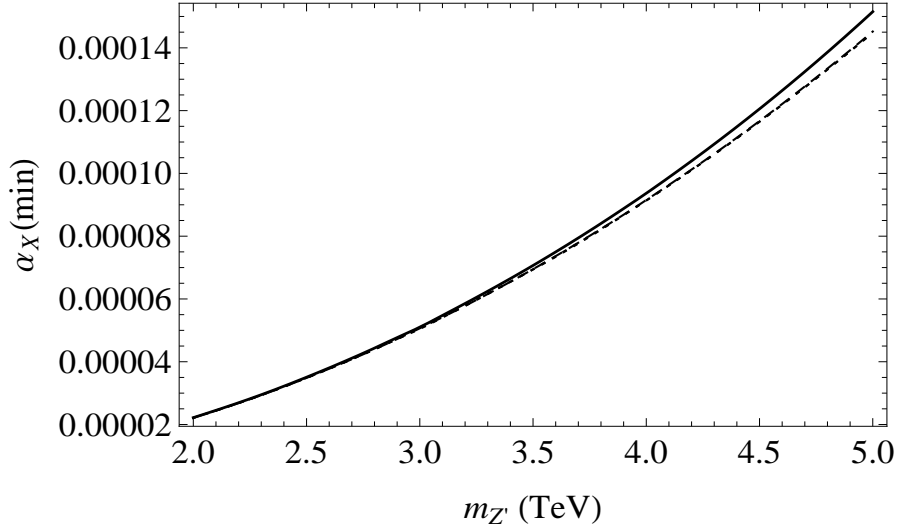} 
\end{center}
\caption{
The lower bounds on $\alpha_X$ as a function of $m_{Z^\prime}$ for various values of $x_H$, 
  in order to reproduce the observed DM relic abundance. 
The solid, dashed, and dotted lines correspond to $x_H=2$, $0$ and $-1.2$, respectively.
The dashed and dotted lines are well overlapping. 
}
\label{fig:alphaVmz}
\end{figure}

In Fig.~\ref{fig:alphaVmz}, we show the lower bound on $\alpha_X$ as a function of 
   $m_{Z^\prime}$ in order to reproduce the DM relic abundance for three fixed values of $x_H$. 
The solid, the dashed, and the dotted lines represent our results for $x_H = 2 $, $0$, and $-1.2$, respectively 
 (the dashed and the dotted lines are well overlapping and indistinguishable). 
The lower bound is increasing as $m_{Z^\prime}$ is raised, since the typical scale of the DM annihilation cross section 
   is controlled by $m_{Z^\prime} \simeq 2 m_{DM}$.  
We can see that the lower bound on $\alpha_X$ is very weakly depends on $x_H$. 
This is because a pair of the RHN DMs dominantly annihilates into the RHNs ($N^{1,2}$) 
   due to their large U(1)$_X$ charges, 
   as indicated by the cross section formulas in Eq.~(\ref{DMSigma}). 

\section{LHC constraints and complementarity with cosmological bounds}
\label{sec:4}
The ATLAS and the CMS collaborations have been searching for a narrow resonance 
  with a variety of final states at the LHC Run-2 with a center-of-mass energy $\sqrt{s}=13$ TeV. 
In the current LHC data, there is no evidence for a resonance state and 
  the upper bound on the resonance productions have been obtained. 
The most severe constraint relevant to the $Z^\prime$ boson in our U(1)$_X$ model 
  is from the resonance search with dilepton final states. 
The latest results by the ATLAS collaboration \cite{ATLAS:2017}  and the CMS collaboration \cite{CMS:2017} 
   with a 36/fb integrated luminosity are consistent with each other, and a lower mass bound of around 4.5 TeV 
   has been obtained for the sequential SM $Z^\prime$ boson. 
In the following analysis, we interpret the current LHC constraints into the $Z^\prime$ boson of our U(1)$_X$ model 
  to obtain an upper bound on U(1)$_X$ gauge coupling as a function of $Z^\prime$ boson mass 
  (for a fixed $x_H$ value).  
Since the ATLAS and CMS results are consistent with each other, we employ the ATLAS result \cite{ATLAS:2017}  
  in our analysis to constrain the model parameters.

The differential cross section for the process, $pp \to Z^\prime +X \to \ell^{+} \ell^{-} +X; \;\ell^{+} \ell^{-}=e^+ e^-/\mu^+ \mu^-$, 
   with respect to the dilepton invariant mass $M_{\ell \ell}$ is given by 
\begin{eqnarray}
 \frac{d \sigma}{d M_{\ell \ell}}
 =  \sum_{q, {\bar q}}
 \int^1_ \frac{M_{\ell \ell}^2}{E_{\rm LHC}^2} dx \, 
  \frac{2 M_{\ell \ell}}{x E_{\rm LHC}^2}  
 f_q(x, Q^2)  \,  f_{\bar q} \left( \frac{M_{\ell \ell}^2}{x E_{\rm LHC}^2}, Q^2
 \right) \times  {\hat \sigma} (q \bar{q} \to Z^\prime \to  \ell^+ \ell^-) ,
\label{CrossLHC}
\end{eqnarray}
where $Q$ is the factorization scale (we fix $Q= m_{Z^\prime}$, for simplicity),  
 $E_{\rm LHC}=13$ TeV is the center-of-mass energy of the LHC Run-2, 
 $f_q$ ($f_{\bar{q}}$) is the parton distribution function for quark (anti-quark), 
  and the cross section for the colliding partons is described as 
\bea 
{\hat \sigma}(q \bar{q} \to Z^\prime \to  \ell^+ \ell^-) =
\frac{\pi}{1296} \alpha_X^2 
\frac{M_{\ell \ell}^2}{(M_{\ell \ell}^2-m_{Z^\prime}^2)^2 + m_{Z^\prime}^2 \Gamma_{Z^\prime}^2} 
F_{q \ell}(x_H),  
\label{CrossLHC2}
\eea
where the function $F_{q \ell}(x_H)$ is given by 
\bea
   F_{u \ell}(x_H) &=&  (8 + 20 x_H + 17 x_H^2)  (8 + 12 x_H + 5 x_H^2),   \nonumber \\
   F_{d \ell}(x_H) &=&  (8 - 4 x_H + 5 x_H^2) (8 + 12 x_H + 5 x_H^2) 
\label{Fql}
\eea
for $q$ being the up-type ($u$) and down-type ($d$) quarks, respectively.
In our analysis, we employ CTEQ6L~\cite{CTEQ} for the parton distribution functions and 
  numerically evaluate the cross section of the dilepton production 
  through the $s$-channel $Z^\prime$ boson exchange. 
Since the RHN DM mass must be close to half of the $Z^\prime$ boson mass, 
  its contribution to the $Z^\prime$ boson decay width is negligibly small, 
  and thus the resultant cross section is controlled by only three free parameters, 
  $\alpha_X$, $m_{Z^\prime}$ and $x_H$.\footnote{
As mentioned before, we have set $m_N^{1,2}=m_{Z^\prime}/4$. 
If we take $m_N^{1,2} \geq m_{Z^\prime}/2$, the RHNs' contribution to the $Z^\prime$ boson 
  decay width is dropped, and hence we obtain the LHC bound as the same as 
  that in the minimal U(1)$_X$ model shown in Refs.~\cite{OO2, Oda:2017kwl, OOR}. 
}    
In interpreting the latest ATLAS results \cite{ATLAS:2017} for the upper bound on 
   the cross section of the process $pp \to Z^\prime +X \to \ell^{+} \ell^{-} +X$, 
   we follow the strategy in Refs.~\cite{OO1, OO2, SO}: we first calculate the cross section of the process 
   by Eq.~(\ref{CrossLHC}) and then we scale our cross section result to find a $k$-factor ($k = 1.31$)  
   by which our cross section coincides with the SM prediction of the cross section presented in the ATLAS paper \cite{ATLAS:2017}. 
This $k$-factor is employed for all of our analysis.   
In this way, we find an upper bound on $\alpha_X$ as a function of $m_{Z^\prime}$ ($x_H$) for a fixed value of $x_H$ 
    ($m_{Z^\prime}$).  
  
The LEP experiments have searched for effective 4-Fermi interactions mediated by a $Z^\prime$ boson \cite{LEPdata}, 
and no significant deviation from the SM predictions have been observed. 
The LEP results are interpreted into a lower bound on $m_{Z^\prime}/\sqrt{\alpha_X}$ for a fixed $x_H$ value,  
  which means an upper bound on $\alpha_X$ as a function of $m_{Z^\prime}$ for a fixed $x_H$ value 
  similar to the constraints obtained from the LHC Run-2 results. 
For the minimal U(1)$_X$ model, the LEP bound on $m_{Z^\prime}/\sqrt{\alpha_X}$ has been obtained in Refs.~\cite{Das:2016zue, OO2}. 
Since the U(1)$_X$ charge assignment for the SM fermions in our model is the same as in the minimal model, 
  the LEP bound presented in Refs.~\cite{Das:2016zue, OO2} can be applied also to our model. 
Thus, we simply refer the bound.    
We will see that the LHC constraints are much more severe than the LEP one for $m_{Z^\prime} \leq 5$ TeV.

To constrain the model parameter space further, we may also consider a theoretical upper bund on $\alpha_X$, 
   namely, the perturbativity bound on the gauge coupling.  
Considering all particles in Table~\ref{table1}, we find the beta function coefficient of the renormalization group (RG) equation
  for the U(1)$_X$ gauge coupling to be 
\bea 
    b_X=\frac{322}{3} + \frac{26}{3} x_H + 7 x_H^2, 
\label{bX}    
\eea 
which is very large in the presence of RHNs and $\Phi_{A, B} $ whose U(1)$_X$ charges are large. 
To keep the running U(1)$_X$ gauge coupling $\alpha_X(\mu)$ in the perturbative regime up to the Planck scale ($M_{Pl}=1.22 \times 10^{19}$ GeV), 
    an upper bound on $\alpha_X$ at low energies can be derived. 
Solving the RG equation for the U(1)$_X$ gauge coupling at the one-loop level, 
  we find the relation between the gauge coupling at $m_{Z^\prime}$ 
  (denoted as $\alpha_X$ in our DM and LHC analysis) and the one at the Planck scale $\alpha_X(M_{Pl})$: 
\bea
  \alpha_{X}  =  \frac{\alpha_{X} (M_{Pl})}{1+ \alpha_{X} (M_{Pl}) \, \frac{ b_X }{2\pi} \ln \left[ \frac{M_{Pl}}{m_{Z^\prime}} \right]}.  
  \label{pert}
\eea 
For simplicity, we have set a common mass for all new particles to be $m_{Z^\prime}$. 
Effects of mass splittings are negligibly small unless new particle mass spectrum is hierarchical. 
Imposing the perturbativity bound of  $\alpha_X(M_{Pl}) \leq 4 \pi$, we find an upper bound on $\alpha_X$ 
   for fixed $m_{Z^\prime}$ and $x_H$ values.

\begin{figure}[t]
\begin{center}
\includegraphics[width=0.49\textwidth, height=5.5cm]{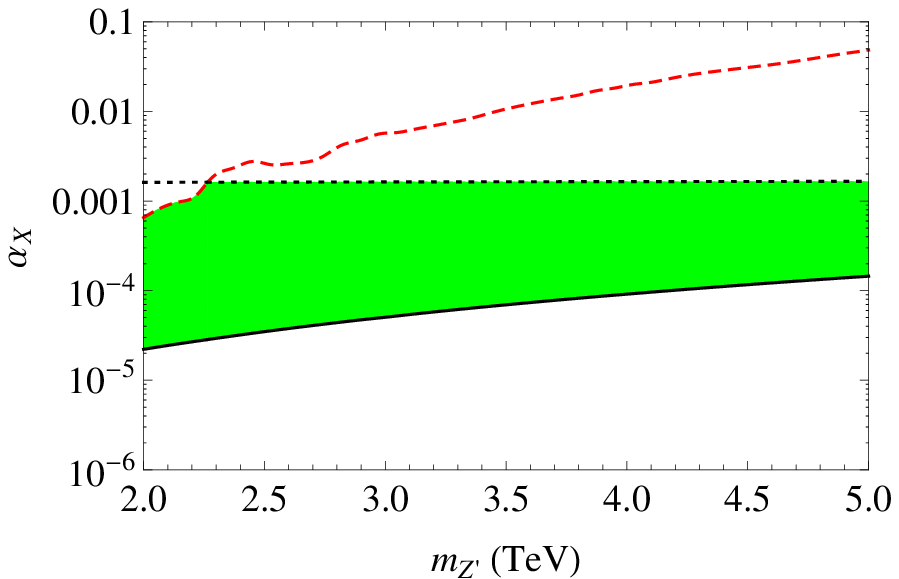}\;
\includegraphics[width=0.49\textwidth, height=5.5cm]{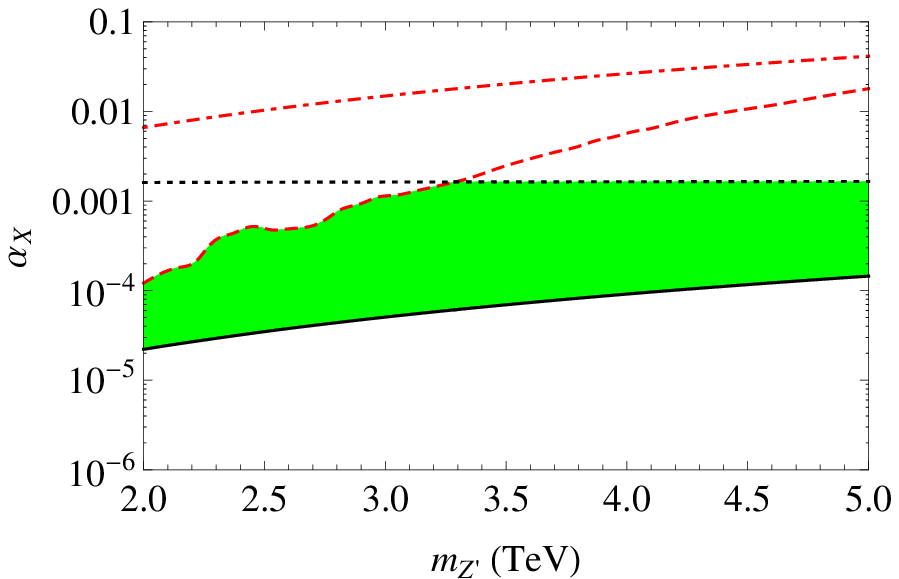}\\ 
\;\;\;\;
\includegraphics[width=0.49\textwidth, height=5.5cm]{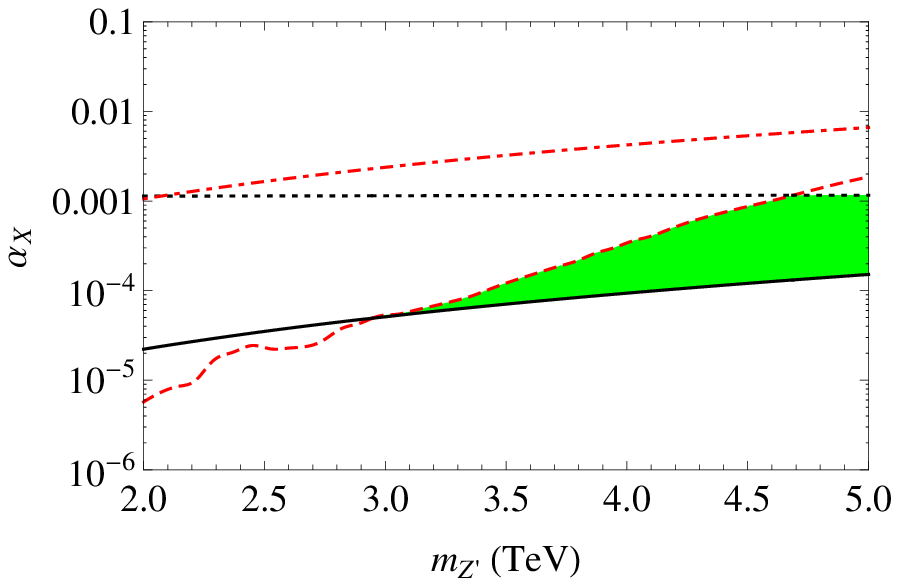}
\end{center}
\caption{
Allowed parameter regions for fixed $x_H$ values in ($\alpha_X, m_{Z^\prime}$)-plane.  
The solid lines are the cosmological lower bounds on $\alpha_X$ as a function of $m_{Z^\prime}$. 
The (red) dashed and (red) dot-dashed lines are the upper bounds on $\alpha_X$ from the LHC and LEP results, respectively. 
The perturbativity bounds on $\alpha_X$ are depicted by the dotted lines. 
The (green) shaded regions satisfy all the constrains. 
The top-left  (right) panel shows the result for a fixed $x_H = -1.2$ ($x_H =0$), and the bottom panel shows the result for $x_H = 2$.
The LEP bound for $x_H = -1.2$ lies outside the range shown in the plot.  
}
\label{fig:LHC}
\end{figure}

Let us now combine all constrains. 
We have obtained the lower bound on $\alpha_X$ from the observed DM relic abundance. 
On the other hand, the upper bound on $\alpha_X$ has been obtained 
  by the LHC results from the search for a narrow resonance, the LEP results 
  and the coupling perturbativity up to the Planck scale. 
Note that these constraints are complementary to narrow down the model parameter space.\footnote{
We see that  the LEP bound is always much weaker than the LHC bounds (for $m_{Z^\prime} \leq 5$ TeV)
  and the perturbativity bound. 
Here, we have considered the LEP bound for completeness.   
} 
In Fig.~\ref{fig:LHC}, we show the combined results for three benchmark $x_H$ values. 
The solid lines are the cosmological lower bounds on $\alpha_X$ as a function of $m_{Z^\prime}$. 
The (red) dashed and (red) dot-dashed lines are the upper bounds on $\alpha_X$ from the LHC and LEP results, respectively. 
The perturbativity bounds on $\alpha_X$ are depicted by the dotted lines. 
The regions satisfying all the constrains are (green) shaded. 
As we have discussed in the previous section, the cosmological lower bound on $\alpha_X$
  depends on $x_H$ very weakly. 
We can see from Eq.~(\ref{bX}) that the perturbativity bound also weakly depends on $x_H$ for $|x_H| < {\cal O}(1)$. 
On the other hand, the LHC bounds are sensitive to $x_H$, and $x_H \simeq -1$ 
  is the best value to loosen the LHC constraints, as we will discuss below. 
For $x_H=-1.2$ (top-left panel), the upper bound on $\alpha_X$ is mainly obtained by the perturbativity bound. 
We can see that as a $x_H$ value is going away from $x_H \simeq -1$, the LHC bounds become more severe. 
For $x_H=2$ (bottom panel), the allowed region is severely constrained to have the lower bound on 
  $m_{Z^\prime} \geq 3$ TeV.

\begin{figure}[t]
\begin{center}
\includegraphics[width=0.75\textwidth, height=7.5cm]{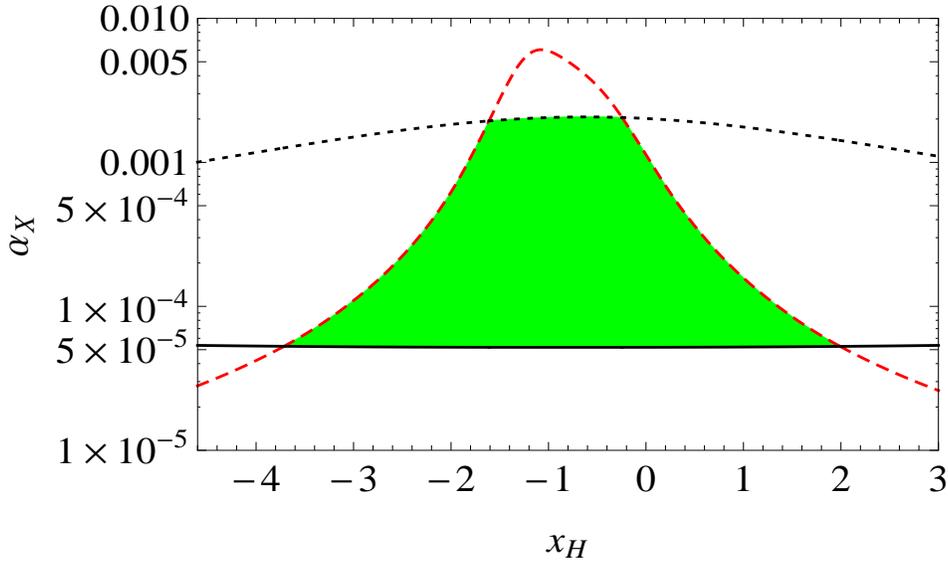} 
\end{center}
\caption{
Allowed parameter region ((green) shaded) for fixed $m_{Z^\prime} = 3$ TeV in ($\alpha_{X}, x_H$)-plane.  
The solid line is the cosmological lower bound on $\alpha_X$ as a function of $x_H$, 
   while the (red) dashed line is the upper bound on $\alpha_X$ obtained from the LHC Run-2 results. 
The dotted line denotes the perturbativity bound. 
The LEP bound lies outside of the range shown in this figure. 
}
\label{fig:alphaVxH}
\end{figure}

Finally, in Fig.~\ref{fig:alphaVxH} we show the allowed parameter region 
   in ($\alpha_{X}, x_H$)-plane for fixed $m_{Z^\prime} = 3$ TeV. 
The solid line is the cosmological lower bound on $\alpha_X$ as a function of $x_H$, 
   while the (red) dashed line is the upper bound on $\alpha_X$ obtained from the LHC Run-2 results. 
The dotted line denotes the perturbativity bound. 
The LEP bound is much weaker than the LHC and the perturbativity bound and lies outside of the range shown in this figure. 
Note again that all three constraints are complementary for narrowing the allowed region (green shaded):   
  $-3.7 \lesssim x_H \lesssim 2.0$ and $5\times 10^{-5} \lesssim \alpha_X \lesssim 2 \times 10^{-3}$. 
The (red) dashed line shows the maximum for $x_H \simeq -1$, 
  which indicates that the dilepton production cross section becomes minimum for this $x_H$ value. 
This fact can be roughly understood by using the narrow decay width approximation. 
When the total decay width of the $Z^\prime$ boson is very narrow, we approximate Eq.~(\ref{CrossLHC2}) as 
\bea 
{\hat \sigma}(q \bar{q} \to Z^\prime \to  \ell^+ \ell^-) \simeq 
\frac{\pi}{1296} \alpha_X^2
M_{\ell \ell}^2 
\left[ \frac{\pi}{m_{Z^\prime} \Gamma_{Z^\prime}} 
\delta( M_{\ell \ell}^2-m_{Z^\prime}^2)
\right]
F_{q \ell}(x_H) 
\propto \frac{F_{q \ell}(x_H)}{F(x_H)+32},   
\eea   
where we have neglected the mass for $N^{1,2}$, for simplicity. 
Using the explicit formulas for $F(x_H)$ and $F_{q \ell}(x_H)$ given in Eqs.~(\ref{F}) and (\ref{Fql}), 
  we can find that the function $F_{q \ell}(x_H)/(F(x_H)+32)$ exhibits a minimum 
  at $x_H=-0.8$ and $x_H \simeq-1.2$ for $q=u$ and $q=d$, respectively. 
The parton distribution functions average the contributions from $u$ and $d$, 
  and we have found that the dilepton production cross section is minimized at $x_H\simeq -1$.

\section{SU(5)$\times$U(1)$_X$ Grand Unification}
\label{sec:5}
In Ref.~\cite{OOR}, the authors of the present paper have proposed 
 SU(5)$\times$U(1)$_X$ grand unification of the minimal U(1)$_X$ model with $Z^\prime$-portal RHN DM. 
In this section, we consider the same direction for our alternative U(1)$_X$ model. 
As discussed in Ref.~\cite{OOR}, the choice of $x_H=-4/5$ allows us 
  to unify the SM gauge group SU(3)$_C \times$SU(2)$_L\times$U(1)$_Y$ into 
  the grand unified SU(5) gauge group \cite{GUT}. 
Although the U(1)$_X$ charge assignment of the RHNs in our model 
  is quite different from the conventional case, 
  the SU(5) grand unification is also possible in our mode by fixing $x_H=-4/5$  
  since the RHNs are singlet under the SM gauge group. 
Interestingly, as shown in Fig.~\ref{fig:alphaVxH}, the choice of $x_H = -4/5$ is close to the best value 
  to result in a wide allowed region. 
This is also true for the SU(5)$\times$U(1)$_X$ unification with the conventional charge assignment \cite{OOR}.
As usual, we consider the SU(5) symmetry breaking by a suitable VEV of a SU(5) adjoint Higgs field 
  with a vanishing U(1)$_X$ charge.

\begin{figure}[t]
\begin{center}
\includegraphics[width=0.49\textwidth, height=5.9cm]{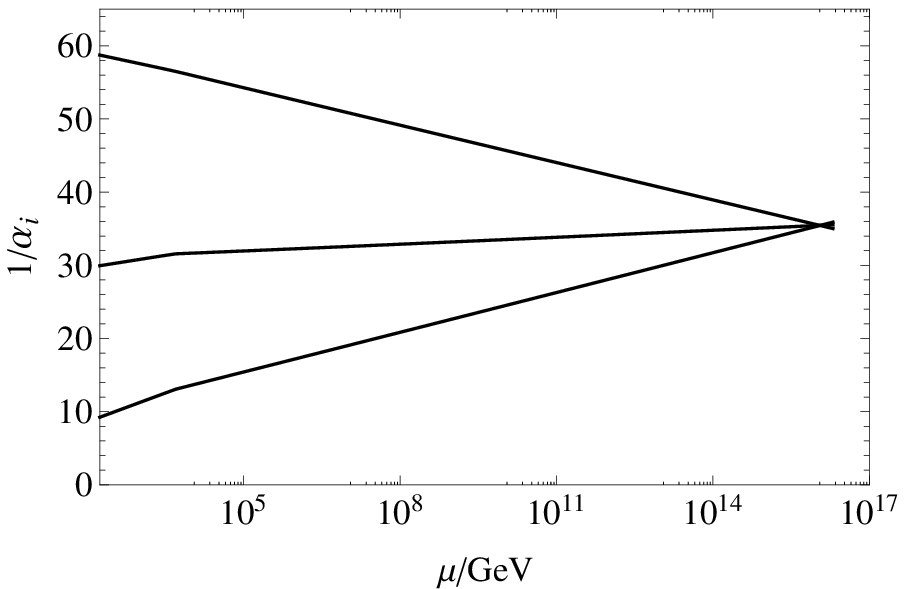}\;
\includegraphics[width=0.49\textwidth, height=5.5cm]{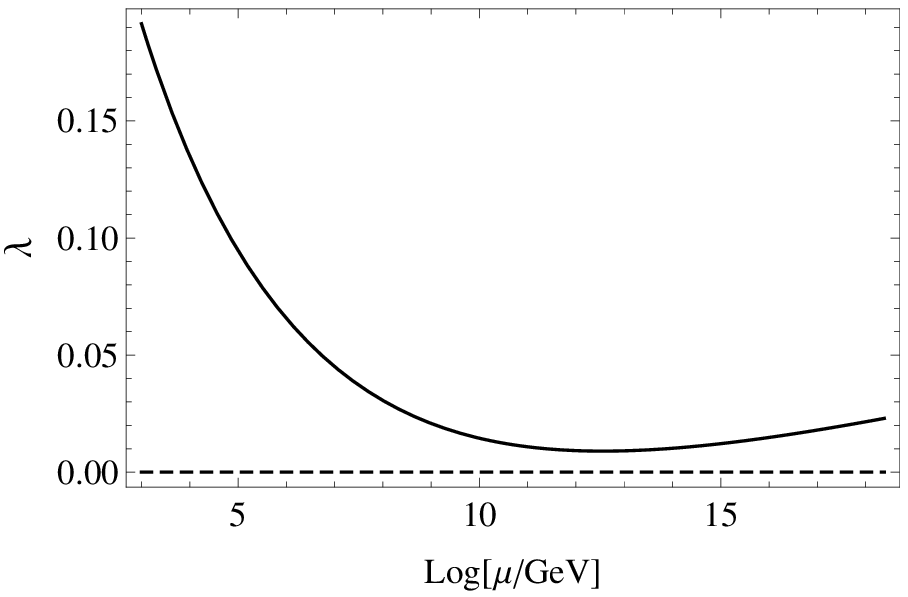}
\end{center}
\caption{
RG running of the SM gauge couplings (left) and the SM Higgs quartic coupling (right).
RG equations employed in our analysis are listed in Appendix. 
}
\label{fig:RGE}
\end{figure}

Let us first consider the unification of the SM gauge couplings.
Following Ref.~\cite{GCU}, we introduce two pairs of vector-like quarks 
   with mass of $\mathcal{O}(1 \; {\rm TeV})$, $D_L +D_R$ and $Q_L +Q_R$ in the representations, 
   $({\bf 3}, {\bf 1}, 1/3)$ and $({\bf 3}, {\bf 2}, 1/6)$ of the SM gauge group, respectively.  
We fix their masses to be larger than $m_{Z^\prime}/2$ not to alter the $Z^\prime$ boson decay width 
   used in the previous sections. 
As have been shown in Ref.~\cite{GCU}, the SM gauge couplings are successfully unified 
  with the SM particle content plus the vector-like quarks at the TeV scale. 
Although our model includes one new Higgs doublet $H_\nu$ being involved in the RG analysis, 
   we have found that the SM gauge couplings are successfully unified at $M_{\rm{GUT}} \simeq 1.13\times 10^{16}$ GeV. 
The RG running of the SM gauge couplings at the one-loop level is shown in the left panel of Fig.~\ref{fig:RGE}.  
In this analysis, we have used a degenerate mass for the vector-like quarks ($M_{D, Q} = 5$ TeV)  
  and a $2.5$ TeV mass for the new Higgs doublet $H_\nu$.

Interestingly, the presence of the vector-like quarks has another phenomenological importance \cite{Gogoladze:2010in}.   
For the observed Higgs boson mass of 125 GeV, it is known that the SM Higgs quartic coupling 
  becomes negative at $\mu \sim 10^{10-11}$ GeV in its RG evolution \cite{Buttazzo:2013uya},  
  and hence the electroweak vacuum is unstable. 
This is because a negative contribution from top quark loops dominates 
   the beta function of the SM Higgs quartic coupling. 
However, the beta function is drastically modified in the presence of the vector-like quarks.
The essential effect is the following: 
The SM SU(2) gauge coupling turns to be asymptotic non-free and becomes larger towards high energies. 
Since the SU(2) gauge coupling yields a positive contribution to the beta function of the SM Higgs quartic coupling, 
   the beta function turns to be positive at some high energy and as a result, the electroweak vacuum instability problem 
   can be solved \cite{Gogoladze:2010in}. 
In the right panel of Fig.~\ref{fig:RGE}, we show the RG evolution of the SM Higgs quartic coupling 
   in the presence of the vector-like quarks with a common mass of $5$ TeV. 
We find that the Higgs quartic coupling is kept positive in its RG evolution.

Under the SU(5)$\times$U(1)$_X$ gauge group, the vector-like quarks, $D_L +D_R^C$ and $Q_L+Q_L^C$, 
  are unified into $({\bf 5}, 3/5) \oplus ({\bf 5}^*, -3/5)$ and $({\bf 10}, 1/5) \oplus ({\bf 10}^*, -1/5)$, respectively. 
To leave only the vector-like quarks light but the others in the multiplets heavy, 
  we consider the same strategy for the triplet-doublet Higgs mass splitting usual in the SU(5) model:  
we introduce their couplings with the SU(5) adjoint Higgs field and tune their mass parameters 
  to make only the vector-like quarks light after the SU(5) gauge symmetry breaking. 
See Ref.~\cite{OOR} for details.

\begin{figure}[t]
\begin{center}
\includegraphics[width=0.75\textwidth, height=8cm]{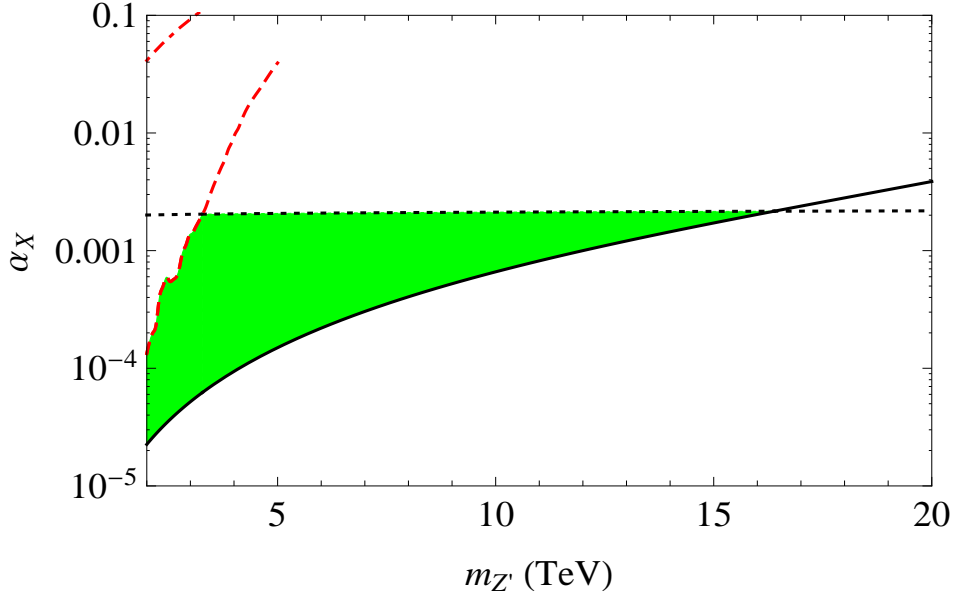}
\end{center}
\caption{
Allowed parameter region ((green) shaded) for the SU(5)$\times$U(1)$_X$ grand unification with $Z^\prime$-portal RHN DM. 
The line codings are same as in Fig.~\ref{fig:LHC}. 
}
\label{fig:LHCSU5}
\end{figure}

Finally, we show the combined result for the SU(5)$\times$U(1)$_X$ grand unification in Fig.~\ref{fig:LHCSU5}. 
Here, we have fixed the vector-like quark masses and the new Higgs doublet mass to achieve 
  the gauge coupling unification, as discussed above.   
Since $x_H=-4/5$ is no longer the free parameter, only two free parameters, $\alpha_X$ and $m_{Z^\prime}$ 
  are involved in this analysis. 
The solid line denotes the lower bound on $\alpha_X$ from the DM relic abundance, 
  while the (red) dashed line and the horizontal dotted line correspond to 
  the upper bound on $\alpha_X$ obtained from the LHC Run-2 constraints and the coupling perturbativity, respectively.  
Here, the perturbativity bound is obtained from Eq.~(\ref{pert}) by imposing $\alpha(M_{\rm GUT}) \leq 4\pi$ 
  with the replacement of $M_{\rm Pl}$ and $b_X$ by $M_{\rm GUT}$ and $b_X= 2666/25$, respectively. 
In Fig.~\ref{fig:LHCSU5}, the LEP constraint (dash-dotted line) is also shown for completeness. 
Combining all the constraints, the resultant allowed parameter region is shown as the (green) shaded region.  
The three constraints are complementary to lead to the upper bound of $m_{Z^\prime} \leq 16.5$ TeV. 
For $m_{DM} \gtrsim 3.5$ TeV, the perturbativity bound on the gauge coupling is more severe than the LHC Run-2 constraint. 
We expect that the High-Luminosity LHC will dramatically improve the constraint,  
  and the (red) dashed line will move to the right in the future to exclude a low $m_{Z^\prime}$ region, 
  otherwise, the evidence of the $Z^\prime$ boson production will be found.

\section{Conclusions and discussions}
\label{sec:Conclusions}

We have considered the non-exotic gauged U(1) extension of the SM. 
Under the new U(1)$_X$ symmetry, the charges of the SM particles are defined 
   as a linear combination of the SM U(1)$_Y$ and U(1)$_{B-L}$ charges.
In the conventional model, three RHNs with a universal U(1)$_X$ charge $-1$ are introduced, 
   with which all the gauge and mixed-gravitational anomalies are canceled.   
In this paper, we have considered an alternative charge assignment for three RHNs, namely, 
  two RHNs ($N_R^{1,2}$)  have a charge $-4$ while one RHN ($N_R$) has a charge $+5$.    
The model with this alternative charge assignment is also free from all the anomalies. 
Introducing a minimal Higgs sector, Majorana neutrino masses for all three RHNs are generated 
  through the spontaneous U(1)$_X$ gauge symmetry breaking. 
Because of the alternative U(1)$_X$ charge assignment, only $N_R^{1,2}$ have 
  Yukawa couplings with the SM lepton doublets, while $N_R$ serves as a unique DM candidate of the model. 
No additional symmetry such as a $Z_2$ parity is necessary to stabilize $N_R$. 
After the electroweak symmetry breaking, the seesaw mechanism generates the SM neutrino mass matrix 
   with only the two Majorana RHNs (minimal seesaw). 

In the context of the alternative U(1) model with the minimal Higgs sector, 
   we have investigated the $Z^\prime$-portal RHN DM scenario. 
Our analysis involves four free parameters, namely, 
   the U(1)$_X$ gauge coupling ($\alpha_X$), 
   a real parameter $x_H$ parametrizing the U(1)$_X$ charge for the SM Higgs doublet,
   the U(1)$_X$ gauge boson ($Z^\prime$) mass ($m_{Z^\prime}$), and  the RHN DM mass ($m_{DM}$).  
We have found that an enhancement of the RHN pair annihilation process via a $Z^\prime$ boson resonance 
  is crucial to reproduce the observed DM relic abundance. 
Thus,  $m_{DM} \simeq m_{Z^\prime}/2$ is required and the freedom of  $m_{DM}$ is effectively dropped out from our analysis. 
We have found a cosmological lower bound on $\alpha_X$ as a function of $m_{Z^\prime}$ ($x_H$) 
   for fixed $x_H$  ($m_{Z^\prime}$) values. 
The current LHC Run-2 results on the search for a narrow resonance with dilepton final states 
   play another important role to constrain our model parameters. 
Employing the LHC Run-2 results, we have found an upper bound 
   on $\alpha_X$ as a function of $m_{Z^\prime}$ ($x_H$) for fixed $x_H$  ($m_{Z^\prime}$) values. 
We have also found that the perturbativity bound on $\alpha_X$ severely constrains the model parameter space.
Combining all the constraints, we have identified an allowed parameter region of the model. 
Here, three constraints from the DM relic abundance, the LHC Run-2 results 
  and the gauge coupling perturbativity are complementary to narrow down the allowed parameter region. 
The present allowed region will be explored by the future collider experiments.

Interestingly, the choice of $x_H = -4/5$, which is very close to the best value to loosen the LHC constraints,
   allows us to extend the model to SU(5)$\times$U(1)$_X$ unification, 
   where the SM gauge group is unified into the grand unified SU(5) group. 
We have shown the SM gauge couplings are successfully unified at $M_{\rm{GUT}} \simeq 1.13\times 10^{16}$ GeV 
   with the introduction of vector-like quarks at the TeV scale. 
With this unification scale, the proton lifetime is estimated as $\tau_p \simeq 6 \times10^{35}$ years, 
   which is consistent with the current experimental lower bound obtained 
   by the Super-Kamiokande \cite{S-K}: $\tau_p (p\to\pi^0 e^+) \gtrsim 10^{34}$ years.  
We have also shown that the SM vacuum instability problem can be solved 
   in the presence of the vector-like quarks at the TeV scale. 
Combining all constraints from the observed DM relic density, the LHC Run-2 results for the $Z^\prime$ boson search,
   and the gauge coupling perturbativity, 
   we have identified a narrow allowed parameter region with the upper bound on $m_{Z^\prime} \leq 16.5$ TeV.    

Finally, we comment on the Higgs sector of our model. 
Because of the alternative U(1)$_X$ charge assignment for three RHNs, the SM Higgs doublet 
  has no coupling with the RHNs in the original Lagrangian, and the neutrino Dirac mass term 
  is generated from the VEV of the new Higgs doublet $H_\nu$. 
This structure is nothing but the one in  the so-called neutrinophilic two Higgs doublet model \cite{nuTHDM}. 
Hence, our model can be regarded as an example of the ultraviolet completion 
   of the neutrinophilic two Higgs doublet model.  
In this model, a very small VEV for $H_\nu$ is induced through 
   a small mixing mass term, $m_{mix}^2 (H^\dagger H_\nu + h.c.)$, 
   and a large positive mass squared, $M^2 H_\nu^\dagger H_\nu$, 
   in the Higgs potential: $v_\nu \sim m_{mix}^2 v/M^2 \ll v$. 
Since such a mixing mass term is forbidden by the U(1)$_X$ gauge symmetry in our model, 
   we may slightly extend our Higgs sector by introducing, for example, a SM singlet Higgs field $\Phi_C$   
   with a U(1)$_X$ charge $-3$. 
Then, the mixing mass term can be induced from a gauge-invariant triple coupling $\Phi_C H^\dagger H_\nu$,
   once $\Phi_C$ develops its VEV. 
Taking all new scalar masses larger than $m_{Z^\prime}/2$, all our results for the DM physics and LHC physics 
   remain intact. 
The contribution of $\Phi_C$ to the beta function coefficient $b_X$ is $3$, which is less than 3\% at most, 
  compared with the total contribution from the other particles. 
Consequently, no significant change emerges for the perturbative bounds we have obtained in the previous sections.

\section*{Acknowledgments}
This work is supported in part by the United States Department of Energy Grant 
No.~DE-SC0013680 (N.O.) and No. DE-SC-0013880 (D.R.), 
and the M. Hildred Blewett Fellowship of the American Physical Society, www.aps.org (S.O.).

\section*{Appendix}
\label{app}
We summarize the RG equations which we have employed for our RG analysis. 
The RG equations for the SM gauge, Yukawa, and Higgs couplings 
 in the presence of a new doublet scalar and two vector-like quarks are given as follows: 
\bea
 \mu  \frac{{\rm d} g_1}{{\rm d}\mu}  = g_1^3 \left( \beta_{g_1}^{\rm 1-loop} +\beta_{g_1}^{\rm 2-loop}\right), \nonumber
\eea
where 
\bea
\beta_{g_1}^{\rm 1-loop}  &=& \frac{1}{16\pi^2} \left(\frac{41}{10} + \frac{2}{5}+  \frac{1}{10} \times \theta(\mu -M_S) \right),
\nonumber \\
\beta_{g_1}^{\rm 2-loop}  &=& \left( \frac{1}{16\pi^2}\right)^2 
\left(
\frac{199}{50} g_1^2+\frac{27}{10} g_2^2+\frac{44}{5} g_3^2 + 
\left(\frac{3}{50} g_1^2+\frac{3}{10} g_2^3+\frac{8}{5} g_3^2\right)\times \theta(\mu -M_Q)
-\frac{17}{10} y_t^2 
\right),
\nonumber 
\eea
and $M_Q$ and $M_S$ are the masses of the vector-like quarks and the new Higgs doublet. 
For simplicity, we set a common mass $M_Q$ for all vector-like quarks. 
\bea
 \mu  \frac{{\rm d} g_2}{{\rm d}\mu}  = g_2^3 \left( \beta_{g_2}^{\rm 1-loop} +\beta_{g_2}^{\rm 2-loop}\right),  \nonumber 
\eea
where 
\bea
\beta_{g_2}^{\rm 1-loop}  &=& \frac{1}{16\pi^2} \left(-\frac{19}{6} + 2 \times \theta(\mu -M_S) +  \frac{1}{6}\right),
\nonumber \\
\beta_{g_2}^{\rm 2-loop} & =& \left( \frac{1}{16\pi^2}\right)^2 
\left(
\frac{9}{10} g_1^2+\frac{35}{6} g_2^2+ 12 g_3^2 
+ \left(\frac{1}{10} g_1^2+\frac{49}{2} g_2^3+8 g_3^2\right)\times \theta(\mu -M_Q) 
-\frac{3}{2} y_t^2 
\right). 
\nonumber 
\eea

\bea
 \mu  \frac{{\rm d} g_3}{{\rm d}\mu}  = g_2^3 \left( \beta_{g_3}^{\rm 1-loop} +\beta_{g_3}^{\rm 2-loop}\right), 
\eea
where 
\bea
\beta_{g_3}^{\rm 1-loop}  &=& \frac{1}{16\pi^2} 
\left(-7 + 2 \times \theta(\mu -M_S) \right), 
\nonumber \\
\beta_{g_3}^{\rm 2-loop}  &=& \left( \frac{1}{16\pi^2}\right)^2 
\left(
\frac{11}{10} g_1^2+\frac{9}{2} g_2^2 - 26 g_3^2 + \left(\frac{1}{5} g_1^2 + 3 g_2^3 + 38 g_3^2\right)\times \theta(\mu -M_Q) - 2 y_t^2 
\right). 
\nonumber 
\eea
\bea
 \mu  \frac{{\rm d} y_t}{{\rm d}\mu}  = y_t \left( \beta_{y_t}^{\rm 1-loop} +\beta_{y_t}^{\rm 2-loop}\right), 
\eea
where 
\bea
\beta_{y_t}^{\rm 1-loop}  &=& \frac{1}{16\pi^2} \left(\frac{9}{2} y_t^2 -\frac{17}{20} g_1^2-\frac{9}{4} g_2^2 - 8 g_3^2
 -\frac{6}{25} g_X^2 \times \theta(\mu -m_{Z^\prime}) \right),
\nonumber \\
\beta_{y_t}^{\rm 2-loop}  &=& \left( \frac{1}{16\pi^2}\right)^2 
\left(
\frac{11}{10} g_1^2+\frac{9}{2} g_2^2 - 26 g_3^2 + \left(\frac{1}{5} g_1^2 + 3 g_2^3 + 38 g_3^2\right) - 2 y_t^2 
\right). 
\nonumber 
\eea

\bea
 \mu  \frac{{\rm d} \lambda}{{\rm d}\mu}  = y_t \left( \beta_{\lambda}^{\rm 1-loop} +\beta_{\lambda}^{\rm 2-loop}\right), 
\eea
where 
\bea
\beta_{\lambda}^{\rm 1-loop}  &=& \frac{1}{16\pi^2} 
\left(
12 \lambda^2 +\left(9 y_t^2 -\frac{9}{5} g_1^2 - 9g_2^2 - \frac{48}{25} g_X^2 \times \theta(\mu -m_{Z^\prime}) \right)\lambda 
\right.
\nonumber \\
&&
\left.
+\frac{9}{4}\left(\frac{3}{25}g_1^4 + \frac{2}{5} g_1^2 g_2^2 +g_2^4 \right)
+ \left(  \frac{72}{125} g_1^2+\frac{24}{25} g_2^2 +\frac{192}{625} g_X^2 
\right) g_X^2 \times \theta(\mu -m_{Z^\prime})  
+12 y_t^2 \lambda -12 y_t^4  
\right),
\nonumber \\
\beta_{\lambda}^{\rm 2-loop}  &=& 
\left(\frac{1}{16\pi^2}\right)^2
\left[
-78\lambda^3
+18\left( \frac{3}{5} g_1^2+3g_2^2\right)\lambda^2
- \left(\frac{73}{8} g_2^4 -\frac{117}{20} g_1^2g_2^2 -\frac{1887}{200} g_1^4\right) \lambda 
\right.
\nonumber \\
 &&\left.-3\lambda y_t^4 +\frac{305}{8} g_2^6
 -\frac{289}{40} g_1^2g_2^4  
 -\frac{1677} {200} g_1^4 g_2^2
 -\frac{3411}{1000} g_1^6
 -64 g_3^2 y_t^4
 -\frac{16}{5} g_1^2y_t^4
 -\frac{9}{2} g_2^4y_t^2 \right.
\nonumber \\
&& \left.+10\lambda \left(\frac{17}{20} g_1^2 +\frac{9}{4} g_2^2 +8g_3^2\right)y_t^2
 -\frac{3}{5}g_1^2 \left(\frac{57}{10}g_1^2 -21g_2^2\right)y_t^2
 -72\lambda^2 y_t^2
 +60y_t^6
 \right]. 
\nonumber 
\eea



\end{document}